\documentclass[usenatbib]{mn2e}
\usepackage{amsmath}
\usepackage{times}
\usepackage{graphicx}
\usepackage[colorlinks=true, linkcolor=blue, citecolor=blue, urlcolor=blue]{hyperref}
\usepackage{xcolor}
\bibpunct{(}{)}{;}{a}{}{,} 
\newcommand{\apj}{ApJ}
\newcommand{\apjl}{ApJL}
\newcommand{\apjs}{ApJS}
\newcommand{\mnras}{MNRAS}

\newcommand{\aap}{A\&A}
\newcommand{\aaps}{A\&AS}
\newcommand{\aj}{AJ}
\newcommand{\araa}{ARA\&A}
\newcommand{\pasj}{PASJ}
\newcommand{\pasp}{PASP}
\newcommand{\na}{NewA}

\begin{document}

\title%
[Dust and gas in NGC\,5485]%
{An extremely low gas-to-dust ratio in the dust-lane lenticular galaxy NGC\,5485\thanks{Herschel is an ESA space observatory with science instruments provided by European-led Principal Investigator consortia and with important participation from NASA.}}

\author%
[M. Baes et al.]{%
Maarten Baes$^1$, 
Flor Allaert$^1$,
Marc Sarzi$^{2,3}$,
Ilse De Looze$^{1,4}$,
Jacopo Fritz$^1$,
\newauthor
Gianfranco Gentile$^{1,5}$,
Thomas M. Hughes$^1$,
Iv\^anio Puerari$^6$,
Matthew W. L. Smith$^7$, 
\newauthor
S\'ebastien Viaene$^1$
\\
$^1$Sterrenkundig Observatorium, Universiteit Gent, Krijgslaan 281, B-9000 Gent, Belgium \\
$^2$Centre for Astrophysics Research, University of Hertfordshire, Hatfield, Hertfordshire, UK \\
$^3$Institut d'Astrophysique de Paris, 98 bis Bd. Arago, 75014 Paris, France\\
$^4$Institute of Astronomy, University of Cambridge, Madingley Road, Cambridge, CB3 0HA, UK \\
$^5$Department of Physics and Astrophysics, Vrije Universiteit Brussel, Pleinlaan 2, 1050, Brussel, Belgium \\
$^6$Instituto Nacional de Astrof\'{\i}sica, Optica, y Electr\'onica, Calle Luis Enrique Erro 1, Santa Mar\'{\i}a Tonantzintla, 72840 Puebla, Mexico \\
$^7$School of Physics and Astronomy, Cardiff University, Queens Buildings, The Parade, Cardiff, CF24 3AA, UK \\
}

\date{Accepted 2014 July 21. Received 2014 July 21; in original form 2014 June 11}
 
\maketitle 

\begin{abstract}
Evidence is mounting that a significant fraction of the early-type galaxy population contains substantial reservoirs of cold interstellar gas and dust. We investigate the gas and dust in NGC\,5485, an early-type galaxy with a prominent minor-axis dust lane. Using new Herschel PACS and SPIRE imaging data, we detect $3.8\times10^6~M_\odot$ of cool interstellar dust in NGC\,5485, which is in stark contrast with the non-detection of the galaxy in sensitive H{\sc{i}} and CO observations from the ATLAS$^{\text{3D}}$ consortium. The resulting gas-to-dust ratio upper limit is $M_{\text{gas}}/M_{\text{d}} < 14.5$, almost an order of magnitude lower than the canonical value for the Milky Way. We scrutinize the reliability of the dust, atomic gas and molecular gas mass estimates, but these do not show systematic uncertainties that can explain the extreme gas-to-dust ratio. Also a warm or hot ionized gas medium does not offer an explanation. A possible scenario could be that NGC\,5485 merged with an SMC-type metal-poor galaxy with a substantial CO-dark molecular gas component and that the bulk of atomic gas was lost during the interaction, but it remains to be investigated whether such a scenario is possible.
\end{abstract}

\begin{keywords}
galaxies: ISM --
galaxies: individual: NGC\,5485
\end{keywords}

\section{Introduction}

Early-type galaxies (ETGs) are often considered as the final stage of galaxy evolution, with virtually little or no cold interstellar medium left. Recently, more and more evidence is appearing that a substantial fraction of the ETGs do contain substantial reservoirs of cold interstellar matter. 

Systematic searches for atomic and molecular gas in ETGs have yielded widely varying detection rates, corresponding to different selection criteria and sensitivity limits \citep[e.g.,][]{2007MNRAS.377.1795C, 2009A&A...498..407G, 2010MNRAS.409..500O, 2010ApJ...725..100W}. The most complete effort to make a census of cold interstellar gas in ETGs in the Local Universe is the ATLAS$^{\text{3D}}$ project \citep{2011MNRAS.413..813C}, which targets a volume-limited sample of 260 ETGs. Molecular gas was detected in 22\% of the galaxies in the sample \citep{2011MNRAS.414..940Y}, whereas the atomic gas detection rate depended strongly on the environment, and increased from 10\% in the Virgo Cluster to 40\% for field ETGs \citep{2012MNRAS.422.1835S}. The frequently observed kinematic misalignments between stars and gas suggests an external origin for the gas in many ETGs \citep[e.g.][]{1993ApJ...419..544S, 2001AJ....121..140K, 2006MNRAS.366.1151S,2008ApJ...676..317Y, 2011MNRAS.417..882D}. Similarly, there is now plenty of evidence for the presence of cold interstellar dust in a significant fraction of the ETG population. Deep optical imaging surveys have shown that many ETGs possess dust features in a variety of morphological forms \citep{1994A&AS..105..341G, 1995AJ....110.2027V, 2001AJ....121.2928T, 2001AJ....121..808C, 2006ApJS..164..334F, 2014NewA...30...51K}. 

In a small fraction of the ETG population, the dust is organized in prominent and large-scale dust lanes \citep{2007A&A...461..103P, 2008MNRAS.390..969F, 2010MNRAS.409..727F}. These so-called dust-lane ETGs are considered to be the remnants of recent minor mergers between ETGs and gas-rich satellites \citep{1981MNRAS.196..747H, 2012MNRAS.423...49K, 2013MNRAS.435.1463K, 2012MNRAS.423...59S}. Combining new ATCA observations and archival data, \citet{2002AJ....123..729O} studied the H{\sc{i}} properties of a sample of 9 dust-lane ETGs. They found huge H{\sc{i}} reservoirs in 6 galaxies, with masses up to several times 10$^9~M_\odot$ and H{\sc{i}}-to-dust mass ratios of more than 1000. On the other side of the spectrum, they did not detect H{\sc{i}} in the remaining 3 dust-lane ETGs, which implied unusually low H{\sc{i}}-to-dust mass ratios, down to $M_{\text{H{\sc{i}}}}/M_{\text{d}} < 20$. They suggested that the cold interstellar medium should mainly be in molecular rather than atomic form in these galaxies. 

\begin{figure}
\centering
\includegraphics[width=\columnwidth]{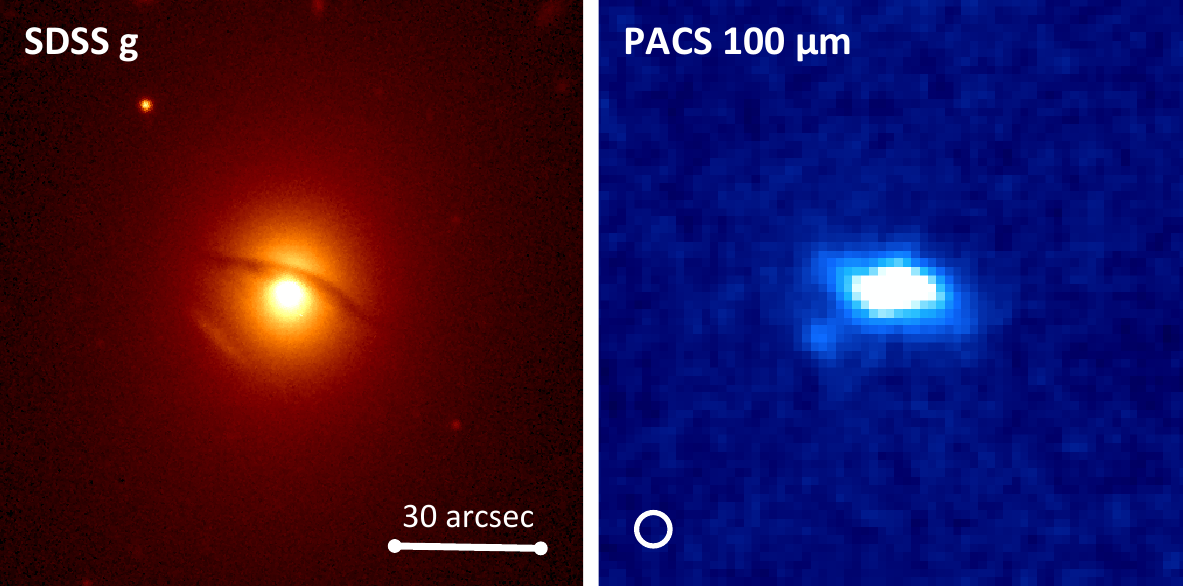}\\[0.5ex]
\includegraphics[width=\columnwidth]{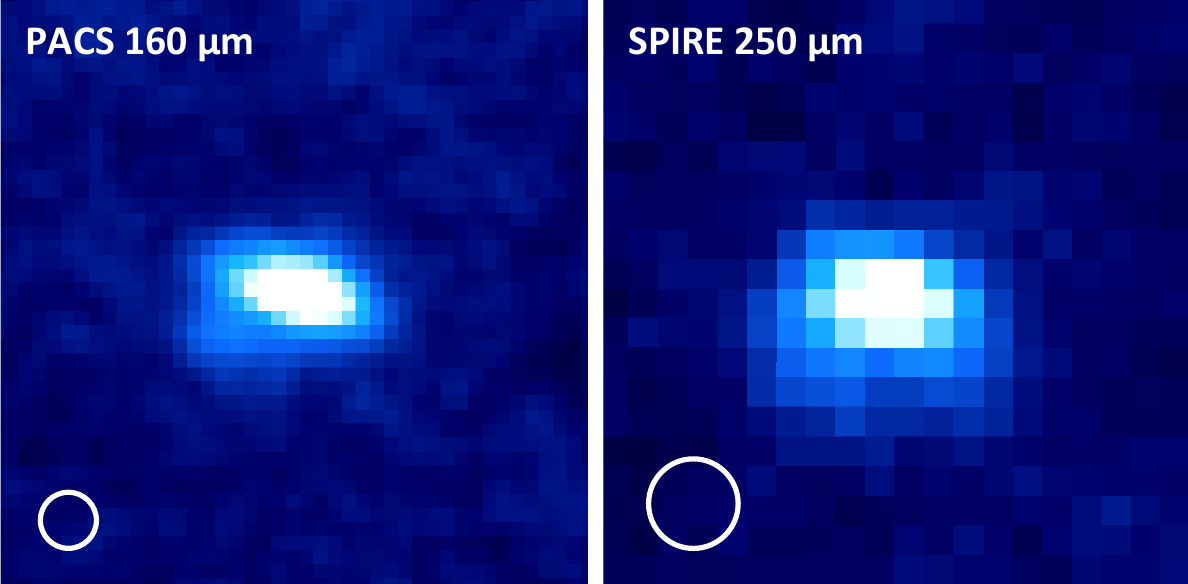}
\caption{SDSS $g$ band, PACS 100 and 160 $\mu$m, and SPIRE 250 $\mu$m images of NGC\,5485. The field of view of each image is $2\times2$ arcmin$^2$, and the beam FWHM is indicated in the three Herschel images. In the optical image, a shell-like feature is clearly visible some 20 arcsec towards the SE of the nucleus, and its counterpart is detectable in the PACS images as well.}
\label{NGC5485-images.pdf}
\end{figure}

In this Letter we focus on NGC\,5485, a dust-lane ETG at a distance of 25.2 Mpc\footnote{Throughout this Letter we use this distance for NGC\,5485; all luminosities and masses are converted to be consistent with this value.} \citep{2001ApJ...546..681T, 2011MNRAS.413..813C}. It is home to a prominent dust lane, perpendicular to the photometric major axis (Fig.~{\ref{NGC5485-images.pdf}}, top left). Given the prominent dust lane, one would expect NGC\,5485 to contain a large cold interstellar matter reservoir. The galaxy is detected by IRAS, with a corresponding dust mass of about $10^6~M_\odot$ \citep{2007A&A...461..103P, 2010MNRAS.407.2475F}. Similarly to a third of the dust-lane ETGs from the \citet{2002AJ....123..729O} sample, NGC\,5485 has a low H{\sc{i}} content: in the frame of the ATLAS$^{\text{3D}}$ campaign, \citet{2012MNRAS.422.1835S} observed the galaxy at 21~cm with the WSRT, but did not detect it. The corresponding upper limit to the atomic gas mass is $M_{\text{H{\sc{i}}}} < 1.5\times10^7~M_\odot$. Surprisingly, also no molecular gas is detected: \citet{2011MNRAS.414..940Y} observed NGC\,5485 in the CO(1-0) and CO(2-1) lines with the IRAM 30m telescope, but did not detect it in either line. The resulting upper limit on the molecular gas mass is $M_{\text{H}_2} < 4.0\times10^7~M_\odot$. 

The combination of these gas and dust mass estimates results in a cold gas-to-dust ratio lower than about 50. This value is probably still strongly overestimated, as IRAS was insensitive to cool dust. In this Letter, we present new far-infrared and submm imaging data for NGC\,5485, taken with the PACS and SPIRE instruments onboard the Herschel Space Observatory. In \S2 we present our observations and the resulting flux densities, and we combine these new data with ancillary data to make a solid measurement of the dust mass, and hence an accurate upper limit on the gas-to-dust ratio. In \S3 we discuss different scenarios to explain this gas-to-dust ratio upper limit, including a potential overestimate of the dust mass, and underestimate of the cold gas mass, and a number of possible alternative options. Finally, \S4 summarizes our conclusions.

\section{Observations and dust mass determination}

NGC\,5485 was observed with the Herschel Space Observatory \citep{2010A&A...518L...1P} in the frame of the {\it{Far-infraRed Investigation of Early-type galaxies with Dust Lanes}} (FRIEDL) program. We used the PACS \citep{2010A&A...518L...2P} and SPIRE \citep{2010A&A...518L...3G} photometers in scan mode, both with their nominal scan speed, i.e. 20 arcsec/s for PACS and 30 arcsec/s for SPIRE. The size of the map was chosen to be $8\times8$~arcmin$^2$. For the PACS map, four cross-scans (i.e. four nominal and four orthogonal scans) were performed, while the SPIRE map was observed with a single cross-scan. The data reduction was done with HIPE version 12.0.0 and includes the same steps as in \citet{2013A&A...556A..54V}. For the PACS data, it includes the use of the Scanamorphos version 23 \citep{2013PASP..125.1126R} in order to make optimal use of the redundancy in the observational data. Figure~{\ref{NGC5485-images.pdf}} shows the resulting Herschel maps at 100, 160 and 250 $\mu$m. At these wavelengths, the FWHM is 7, 11 and 18 arcsec respectively, and the dust emission from the galaxy is resolved along the direction of the dust lane. At 350 and 500 $\mu$m, where the FWHM increases to 24 and 36 arcsec, the emission is not resolved anymore.

Global flux densities were determined through aperture photometry using the DS9/Funtools program FUNCTS, following the same strategy as described in detail by \citet{2013A&A...556A..54V}. We followed the approach of \citet{2012ApJ...745...95D} to determine the background level and uncertainty estimates. The Herschel flux densities are $533\pm 44$, $797\pm52$, $518\pm46$, $288\pm33$ and $108\pm23$ mJy at 100, 160, 250, 350 and 500 $\mu$m.

To determine the dust mass, we first used a simple modified blackbody model, with dust emissivity index $\beta=2$ and $\kappa_\nu=0.192$~m$^2$\,kg$^{-1}$ at 350~$\mu$m. These values correspond to what is probably the most widely adopted physical dust model \citep{2003ARA&A..41..241D, 2007ApJ...657..810D}, and they have been widely used to interpret Herschel SEDs \citep[e.g.,][]{2012ApJ...745...95D, 2013MNRAS.428.1880A, 2013A&A...556A..54V, 2014A&A...565A...4H}. Applying such a simple modified blackbody fit to the PACS and SPIRE flux densities, we find $T_{\text{d}} = 18.8\pm0.4$~K and $M_{\text{d}} = (3.79\pm0.31)\times10^6~M_\odot$.

\begin{figure}
\centering
\includegraphics[width=\columnwidth]{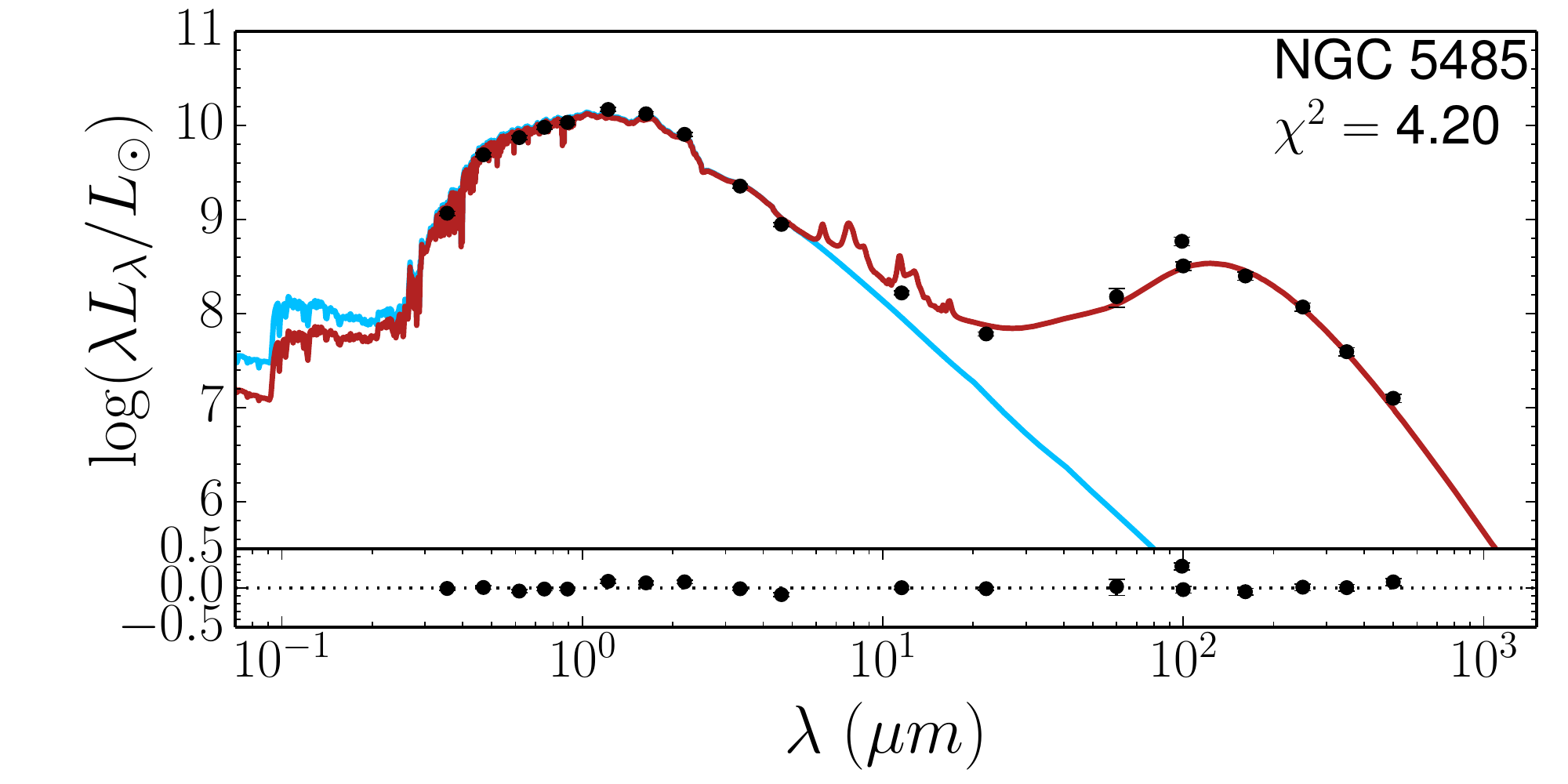}
\caption{A MAGPHYS fit to the UV-submm spectral energy distribution of NGC\,5485. The top panel shows the SED, where the red stars represent the SDSS, 2MASS, WISE, IRAS and Herschel data points, the blue solid line is the intrinsic stellar SED, and the solid black line is the model SED including dust extinction and emission. The bottom panel shows the residuals between the data and the model. The $\chi^2$ value is dominated by the IRAS 100 $\mu$m flux density, which is inconsistent with the PACS 100 $\mu$m flux density.}
\label{NGC5485-MagPhys.pdf}
\end{figure}

Apart from a simple modified blackbody fit, we also used the MAGPHYS code \citep{2008MNRAS.388.1595D} to determine the dust mass. MAGPHYS has been used extensively to model the panchromatic SEDs of galaxies \citep[e.g.,][]{2012MNRAS.427..703S, 2013ApJ...768...90L, 2013MNRAS.433..695C, 2014A&A...567A..71V}. We combined the Herschel PACS and SPIRE flux densities with optical {\it{ugriz}} flux densities for NGC\,5485 from SDSS, near-infrared {\it{JHK}}$_{\text{s}}$ flux densities from 2MASS, mid-infrared flux densities from WISE, and far-infrared flux densities at 60 and 100 $\mu$m from IRAS. The result of the MAGPHYS fit can be found in Figure~{\ref{NGC5485-MagPhys.pdf}}. The most important property for our goals is the total dust mass, for which we found $M_{\text{d}} = (3.81^{+0.80}_{-0.64})\times10^6~M_\odot$, completely consistent with the value obtained from our modified blackbody fit (we adopted the same dust model).

\section{Discussion}

Combining the dust mass resulting from the Herschel flux densities with the molecular and atomic gas measurements discussed in \S1, we find a cold gas-to-dust upper limit $M_{\text{gas}}/M_{\text{d}} < 14.5$. Such an gas-to-dust ratio is extremely low -- typical values range from several thousands for low-metallicity dwarf galaxies to about 100 for solar metallicity galaxies \citep[e.g.,][]{2009ApJ...701.1965M, 2011A&A...532A..56G, 2011A&A...535A..13M, 2014A&A...563A..31R}.

One possible explanation for the extreme gas-to-dust ratio upper limit for NGC\,5485 is that the dust mass as determined from the Herschel flux densities is overestimated. There are several possible systematic effects that we should investigate in some detail.

A first observation is that the new dust mass we have derived is a factor 3 to 4 larger than the IRAS-based dust masses of \citet{2007A&A...461..103P} and \citet{2010MNRAS.407.2475F}. And these dust masses are themselves more than an order higher than the dust masses the same authors derived using extinction in optical images, for which they found several times $10^4~M_\odot$. However, we are convinced that our new dust mass based on Herschel FIR/submm data is more reliable than those based on IRAS and optical extinction. IRAS-based dust masses are most certainly an underestimate of the true dust mass, since IRAS is insensitive to cool dust at temperature below about 25~K \citep[e.g.,][]{1995A&A...298..784G, 1996ApJ...468..571T}. Herschel observations have decisively shown that the typical cold dust temperature in ETGs is around 20~K \citep{2012ApJ...748..123S, 2012MNRAS.419.2545R, 2013A&A...552A...8D}. For NGC\,5485, we find an effective dust temperature of 19~K, which indeed indicates that IRAS missed the bulk of the dust. On the other hand, inferring the amount of dust from optical extinction maps is complicated due to the complexity of the star-dust geometry and the effects of scattering \citep[e.g.,][]{1992ApJ...393..611W, 2001MNRAS.326..733B}. In dust-lane ETGs, dust mass estimates based on simple extinction recipes applied to optical images can underestimate the true dust mass by an order of magnitude or more \citep{1995A&A...298..784G, 2007A&A...461..103P, 2008MNRAS.390..969F, 2010MNRAS.409..727F}.

We should also critically consider the possibility that the observed Herschel flux densities are overestimated. NGC\,5485 has a galactic latitude of 59$^\circ$, so there are no foreground subtraction issues due to strong cirrus that have plagued some other Herschel extragalactic observations \citep[e.g.,][]{2010MNRAS.409..102D, 2012A&A...546A..34F}. In optical and NIR images, NGC\,5485 seems to have a shell-like structure some 20 arcsec towards to SE of the nucleus (see left panel of Figure~{\ref{NGC5485-images.pdf}}). This feature has a clear counterpart in the PACS maps, and could be an edge-on background galaxy rather than a feature linked to NGC\,5485 itself. In any case, the contribution of this source to the Herschel flux densities of NGC\,5485 is marginal. In the PACS 100 $\mu$m map, where the resolution is sufficient to separate this feature from the dust emission in the dust lane, we find that it contributes no more than 10\% to the total flux density. 

Finally, a last aspect to take into consideration is that we modeled the FIR/submm emission as a single modified blackbody at a single temperature, whereas in reality, the FIR/submm emission results from a complicated weighted sum of modified blackbodies at a range of temperatures. This is a systematic effect that we cannot avoid due to our lack of resolution. However, the total dust mass is typically {\em{underestimated}} when it is based on the integrated SED \citep{2011A&A...536A..88G, 2012MNRAS.425..763G}, so it does not help to explain the low gas-to-dust ratio.

On the other hand, the low gas-to-dust ratio might be the result of a systematic underestimate of the gas mass. In any case, there does not seem to be any reason to doubt the reliability of the H{\sc{i}} and CO upper limits of \citet{2012MNRAS.422.1835S} and \citet{2011MNRAS.414..940Y}, respectively. The noise levels obtained in both observation campaigns, with the WSRT and IRAM 30m telescope respectively, are almost exactly the median value of the entire ATLAS$^{\text{3D}}$ study. One source of uncertainty on the derivation of the molecular gas mass upper limit is the conversion factor needed to convert CO(1-0) intensities to the H$_2$ masses, the so-called $X_{\text{CO}}$ factor. This factor has been the subject of a substantial debate in the recent literature \citep[e.g.,][]{1996A&A...308L..21S, 1997A&A...328..471I, 2011ApJ...737...12L, 2012ApJ...756...40S, 2013ApJ...777....5S}. The authoritative review by \citet{2013ARA&A..51..207B} suggests the value $X_{\text{CO}} = 2\times10^{20}$ cm$^{-2}$ (K~km~s$^{-1}$)$^{-1}$ for the Milky Way and other "normal galaxies". In their determination of the upper limit for NGC\,5485, \citet{2011MNRAS.414..940Y} used $X_{\text{CO}} = 3\times10^{20}$ cm$^{-2}$ (K~km~s$^{-1}$)$^{-1}$, i.e.\ 50\% larger than the standard value suggested by \citet{2013ARA&A..51..207B}. Using the standard value would decrease the molecular gas mass upper limit by factor 1.5, and it would decrease the gas-to-dust ratio upper limit by a factor 1.32 to $M_{\text{gas}}/M_{\text{d}} < 11$.

In principle, most of the gas in NGC\,5485 could be ionized rather than neutral. An interesting comparison case here is the study of the elliptical galaxy NGC\,4125 by \citet{2013ApJ...776L..30W}. Combining Herschel-based dust estimates with H{\sc{i}} and CO non-detections, they also found a similar cold gas-to-dust ratio upper limit as we obtain here. However, based on a [N{\sc{ii}}]/[C{\sc{ii}}] ratio consistent with a low-density ionized medium and the (crude) assumption that the ionized gas is distributed uniformly over a 4.2 kpc diameter sphere, \citet{2013ApJ...776L..30W} argued that a significant fraction of the gas in NGC\,4125 is warm ionized rather than cold neutral gas. For NGC\,5485, this option seems excluded, though. The galaxy was imaged in H$\alpha$ by \citet{2010MNRAS.407.2475F}. They detected an inclined disc of ionized gas that nicely follows the morphology of the dust lane, which seems to support that at least part of the dust might be associated with warm ionized gas. The total amount of ionized hydrogen gas traced in the H$\alpha$ map is $M_{\text{H{\sc{ii}}}} = (2.0\pm0.6)\times10^4~M_\odot$, far too little to be a significant contributor to the total gas budget. \citet{2010MNRAS.407.2475F} even warn that their H{\sc{ii}} masses are likely overestimated by a factor 2-3 due to confusion of the H$\alpha$ and [N{\sc{ii}}] lines. On the other hand, H$\alpha$ photons are subject to dust attenuation in the dust lane, which could cause the H{\sc{ii}} mass in NGC\,5485 to be underestimated. However, in order to bring the H{\sc{ii}} mass to the level required for a typical gas-to-dust ratio of about 100, unrealistically high attenuation values $A_{\text{V}}\gg10$ are needed \citep{1994ApJ...429..582C}.

Alternatively, a major part of the gas budget in NGC\,5484 could be in the form of hot X-ray emitting ionized gas. While dust grains are expected to be destroyed through sputtering by thermal collisions with energetic ions, the coexistence of dust and a hot X-ray gas is not completely impossible \citep[e.g.,][]{2003ApJ...585L.121T, 2007PASJ...59..107K, 2008ApJ...677..249L}. NGC\,5485, however, was not detected in the ROSAT all-sky survey: \citet{1999MNRAS.302..209B} quote an upper limit of $L_{\text{X}} < 10^{39.8}~{\text{erg}}\,{\text{s}}^{-1}$. If we insert the K-band luminosity $L_{\text{K}} = 5.65\times10^{10}~L_\odot$ into the latest calibration between the K-band luminosity of ETGs and the expected unresolved X-ray emission due to low-mass X-ray binaries (LMXBs) by \citet{2011ApJ...729...12B}, we find exactly the same number, $L_{\text{X,LXMB}} = 10^{39.8}~{\text{erg}}\,{\text{s}}^{-1}$. This implies that any X-ray emission from NGC\,5485 most probably originates from LMXBs rather than from hot ionized gas. Moreover, the fact that  the dust is distributed in a well-defined dust lane makes a physical association with a potential hot diffuse halo rather unlikely, although it is not impossible that a fraction of the dust is distributed diffusely over the galaxy \citep{1995A&A...298..784G}.

A final, more exotic, option could be that NGC\,5485 contains a reservoir of molecular gas, but that this gas is not emitting CO line emission. Both diffuse  $\gamma$-ray emission \citep{2005Sci...307.1292G, 2010ApJ...710..133A, 2012ApJ...750....3A} and combined gas and dust observations \citep{2011A&A...536A..19P, 2012A&A...543A.103P} have revealed the presence of a substantial amount of so-called dark gas in our own Milky Way. The presence of CO-dark gas has been found many years ago in low-metallicity dwarf galaxies. At low metallicities, most of the carbon is locked up in neutral or singly ionized carbon rather than CO, which makes [C{\sc{ii}}] a more reliable tracer the cold H$_2$ reservoir \citep{1991ApJ...373..423S, 2010ApJ...724..957S, 1997ApJ...483..200M, 2013PASP..125..600M}. As a relatively massive lenticular galaxy, NGC\,5485 is not expected to contain copious amounts of CO-dark gas. Being a dust-lane ETG, however, it has most probably acquired most of its dust and gas during a recent minor merger \citep{2012MNRAS.423...49K}. The recent merger scenario is also supported by its rather exceptional kinematical structure, which shows strong minor-axis rotation \citep{1988A&A...195L...5W, 2011MNRAS.414.2923K, 2011MNRAS.414..888E}. 

In principle, it is possible that it has accreted a dwarf galaxy with a substantial CO-dark molecular gas reservoir. An appropriate example of such a dwarf galaxy would be the SMC, which contains a dust mass of $3\times10^5~M_\odot$, but hardly any CO-emitting molecular gas \citep{2001PASJ...53L..45M, 2007ApJ...658.1027L, 2011ApJ...737...12L}. It does, however, contain about $4\times10^8~M_\odot$ of H{\sc{i}} gas \citep{1999MNRAS.302..417S, 2011ApJ...741...12B}, so if a minor merger with an SMC-type galaxy would be the origin of the dust lane in NGC\,5485, somehow a large fraction of this atomic gas should have been lost during the merger event. Atomic gas is generally more loosely bound than molecular gas, so it could be more easily be stripped during interactions (but it would probably still be visible as a tidal tail). Another argument against such a scenario is that the gas-to-dust ratios of low-metallicity dwarf galaxies are typically up to an order of magnitude higher than those of giant galaxies \citep{2014A&A...563A..31R}, which makes the apparently low gas-to-dust ratio in NGC\,5485 even more puzzling. Whether such a merging scenario is plausible or even possible, is a challenge that could be investigated using detailed merger hydrodynamical simulations. Also deeper H{\sc{i}} and CO observations, sensitive enough to trace possible diffuse gas, would be useful to unveil the nature of the ISM in this peculiar system. 

\section{Conclusions}

We have discussed the interstellar dust and gas properties of the dust-lane ETG NGC\,5485. We present new Herschel PACS and SPIRE imaging of NGC\,5485, taken in the frame of the {\it{Far-infraRed Investigation of Early-type galaxies with Dust Lanes}} (FRIEDL) program. Using both standard modified blackbody model fits and the MAGPHYS spectral energy distribution modeling code, we obtain a dust mass $M_{\text{d}} = 3.8\times10^6~M_\odot$. The combination of the dust mass and H{\sc{i}} and CO stringent upper limits, obtained in the frame of the ATLAS$^{\text{3D}}$ survey, leads to an exceptionally low gas-to-dust ratio, $M_{\text{gas}}/M_{\text{d}} < 14.5$, almost an order of magnitude lower than the canonical value of the Milky Way. 

We have investigated different possible explanations for this extreme gas-to-dust ratio. We have critically checked the reliability of the dust mass estimate, but neither the lack of spatial resolution in the FIR/submm imaging, nor a possible contamination of background or companion sources affect the dust mass. Similarly, the reliability of the cold gas mass is scrutinized. The main source of uncertainty here is the notorious CO-to-H$_2$ conversion factor, and if we would assume the "standard" value rather than the slightly higher value adopted by the ATLAS$^{\text{3D}}$ consortium, the gas-to-dust ratio upper limit would even by decreased to $M_{\text{gas}}/M_{\text{d}} < 11.0$.

Finally, we investigate less obvious scenarios to explain the lack of cold gas in NGC\,5485. Based on H$\alpha$ and X-ray observations, we can discard the possibility that the bulk of the gas is in a warm of hot ionized medium. One possible option is the presence of a component of CO-dark molecular gas. In principle, the extreme gas-to-dust ratio could be the result of a merger with an SMC-type metal-poor dwarf galaxy, if a substantial fraction of the H{\sc{i}} could have been lost during the interaction, but it remains to be investigated whether such a scenario is possible.

\section*{Acknowledgements}

The authors are indebted to the anonymous referee for providing insightful and constructive comments. MB, JF and TH acknowledge the financial support of the Belgian Science Policy Office (BELSPO) through the PRODEX project "Herschel-PACS Guaranteed Time and Open Time Programs: Science Exploitation" (C90370). MB, FA, SV and IDL acknowledge the support from the Flemish Fund for Scientific Research (FWO-Vlaanderen). IP thanks the Mexican foundation CONACyT for financial support.

PACS has been developed by a consortium of institutes led by MPE (Germany) and including UVIE (Austria); KU Leuven, CSL, IMEC (Belgium); CEA, LAM (France); MPIA (Germany); INAF-IFSI/OAA/OAP/OAT, LENS, SISSA (Italy); IAC (Spain). This development has been supported by the funding agencies BMVIT (Austria), ESA-PRODEX (Belgium), CEA/CNES (France), DLR (Germany), ASI/INAF (Italy), and CICYT/MCYT (Spain). SPIRE has been developed by a consortium of institutes led by Cardiff University (UK) and including Univ. Lethbridge (Canada); NAOC (China); CEA, LAM (France); IFSI, Univ.\ Padua (Italy); IAC (Spain); Stockholm Observatory (Sweden); Imperial College London, RAL, UCL-MSSL, UKATC, Univ.\ Sussex (UK); and Caltech, JPL, NHSC, Univ.\ Colorado (USA). This development has been supported by national funding agencies: CSA (Canada); NAOC (China); CEA, CNES, CNRS (France); ASI (Italy); MCINN (Spain); SNSB (Sweden); STFC (UK); and NASA (USA).


\bibliography{NGC5485}

\end{document}